\documentstyle[12pt,epsf]{article}

\def\beq{\begin{equation}}
\def\eeq{\end{equation}}
\def\beqar{\begin{eqnarray}}
\def\eeqar{\end{eqnarray}}

\def\he#1{\hbox{${}^{#1}$He}}
\def\li#1{\hbox{${}^{#1}$Li}}

\def\yp{\hbox{$Y_{\rm p}$}}

\def\la{\mathrel{\mathpalette\fun <}}
\def\ga{\mathrel{\mathpalette\fun >}}
\def\fun#1#2{\lower3.6pt\vbox{\baselineskip0pt\lineskip.9pt
  \ialign{$\mathsurround=0pt#1\hfil##\hfil$\crcr#2\crcr\sim\crcr}}}

\begin{document}

\def\lsim{\mathrel{\vcenter{\hbox{$<$}\nointerlineskip\hbox{$\sim$}}}}
\def\gsim{\mathrel{\vcenter{\hbox{$>$}\nointerlineskip\hbox{$\sim$}}}}
\def\erf{\mathop{\rm erf}\nolimits}
\def\nnu{N_\nu}
\def\Hunit{km s$^{-1}$ Mpc$^{-1}$}
\def\ie{i.e.}
\def\eg{e.g.}
\def\etal{et al.}
\def\PLB{Phys.\ Lett.\ B}
\def\PRC{Phys.\ Rev.\ C}
\def\PRD{Phys.\ Rev.\ D}
\def\PRL{Phys.\ Rev.\ Lett.}
\def\MNRAS{Mon. Not. R. Astr. Soc.\ }
\def\ApJ{Ap.\ J.}
\def\ApJS{Ap.\ J.\ Suppl.\ Ser.}

\begin{titlepage}
\pagestyle{empty}
\baselineskip=21pt
\rightline{UMN-TH-1515/96}
\rightline{hep-ph/9610319}
\rightline{October 1996}
\vskip .2in
\baselineskip=15pt
\bigskip

\begin{center}
{\bf A BIG BANG NUCLEOSYNTHESIS LIKELIHOOD ANALYSIS \\
OF THE BARYON-TO-PHOTON RATIO AND \\
THE NUMBER OF LIGHT PARTICLE DEGREES OF FREEDOM}
\end{center}

\bigskip
\centerline{Keith A. Olive\footnote{olive@mnhep.hep.umn.edu}
}
\smallskip
\centerline{\it University of Minnesota, School of Physics and Astronomy}
\centerline{\it 116 Church St SE, Minneapolis, MN 55455}

\bigskip
\centerline{David Thomas\footnote{davet@phys.ufl.edu}
}
\smallskip
\centerline{\it University of Florida, Dept.\ of Physics}
\centerline{\it Box 118440, Gainesville, FL 32611-8440}
\bigskip

\centerline{\bf Abstract}

{\narrower
\noindent
We extend the model-independent likelihood analysis of 
big bang nucleosynthesis (BBN) based on \he4 and \li7 to allow for 
numbers of degrees of freedom which differ from the standard model
value characterized by $N_\nu = 3$. 
We use the two-dimensional likelihood functions to simultaneously
constrain the baryon-to-photon ratio and
the number of light neutrinos.  The upper limit thus obtained is
$\nnu<4.0$ (at 95\% C.L.).
We also consider the consequences
if recent observations of deuterium in high-redshift QSO absorption-line
systems (QSOALS) are confirmed.
\par}
\end{titlepage}
\vfill\eject
\baselineskip=\normalbaselineskip

\section{Introduction}

In recent years the comparison between standard big bang nucleosynthesis 
and observational data on elemental abundances has come under close 
scrutiny.  Minor changes in the theoretical abundances (improvements 
in the code, updated neutron lifetime data) combined with 
many new determinations of the abundances of the light element
isotopes, D, \he3, \he4, and \li7 have led to a flurry of sophisticated
statistical analyses of BBN \cite{kr,skm,kk1,kk2,hata1,hata2,fo,fkot}.  
The results of these analyses however 
depend strongly on the assumptions made about galactic chemical
evolution, and in particular, the chemical evolution 
of deuterium.  For the most part, the comparison between BBN theory and 
the observational data made heavy use of the solar value of D + \he3
because 
it  provided a (relatively large) lower limit for the baryon-to-photon
ratio \cite{ytsso}, $\eta_{10} \equiv 10^{10}\eta > 2.8$.
This limit for a long time was seen to be
essential because it provided the only means for bounding $\eta$ from below
and in effect allows one to set an upper limit on the number of neutrino
flavors \cite{ssg}, $N_\nu$, as well as other constraints on particle physics
properties. The upper bound to $N_\nu$ 
is in fact strongly dependent on the lower bound to
$\eta$.  For lower $\eta$ the \he4 abundance predicted by BBN drops,
allowing for a larger $N_\nu$ to match a given primordial \he4 abundance,
\yp, as determined from the observations.
For $\eta < 4 \times 10^{-11}$,
there is no
bound whatsoever on $N_\nu$ \cite{ossty}. 

It was argued \cite{ytsso} that since stars (even massive stars) do not 
destroy \he3 in its entirety, we can obtain a bound on $\eta$ from an
upper bound to the solar D and \he3 abundances. One can in fact limit
\cite{ytsso,ped}
 the sum of primordial D and \he3 by applying the expression below at $t =
\odot$
\beq
{\rm \left({D + \he3 \over H} \right)_p \le \left({D \over H} \right)_t}
+ {1 \over g_3}{\rm  \left({\he3 \over H} \right)_t}
\label{he3lim}
\eeq
In (\ref{he3lim}), $g_3$ is the fraction of a star's initial D and \he3 which
survives as \he3. For $g_3 > 0.25$ as suggested by standard stellar models, 
and using the
solar data on D/H and
\he3/H, one finds $\eta_{10} > 2.8$. 
This argument has been improved
recently \cite{st} ultimately leading to a stronger limit \cite{hata2} 
$\eta_{10} > 3.8$ and a best estimate $\eta_{10} = 6.6 \pm 1.4$.
The stochastic approach used in  \cite{cst} could
only lower the bound from 3.8 to about 3.5 when assuming as always that $g_3 >
0.25$.

The limit $\eta_{10} > 2.8$ derived using (\ref{he3lim}) is really a one
shot approximation.  Namely, it is assumed that material passes 
through a star no
more than once. However, to determine whether or not the solar (and present) 
values of D/H and \he3/H can be matched to a given primordial abundance,
it is necessary to consider models of galactic chemical
evolution. In the absence of stellar \he3 production, 
particularly by
low mass stars, it was shown \cite{vop} that there are indeed suitable choices
for a star formation rate and an initial mass function to: 1) match the D/H
evolution from a primordial value (D/H)$_{\rm p} = 7.5 \times 10^{-5}$,
corresponding to $\eta_{10} = 3$, through the solar and present
interstellar medium (ISM) abundances,  while 2)
at the same time keeping the \he3/H evolution relatively flat so as not to
overproduce \he3 at the solar and present epochs. This was achieved for $g_3 >
0.3$.  

In the context of models of galactic chemical evolution, there is, however, 
 little justification a
priori for neglecting the production of \he3 in low mass
stars. Indeed, stellar models predict that considerable amounts of \he3 are
produced in stars between 1 and 3 M$_\odot$ \cite{it} and this prediction is 
consistent with the observation of high \he3/H 
in planetary nebulae \cite{rood}.
Generally, implementation of these \he3 yields in chemical
evolution models leads to a gross overproduction of \he3/H particularly at the
solar epoch \cite{orstv}.

As indicated earlier, 
the presence of a lower bound on $\eta$ allows us to place an upper
bound to the number of neutrino flavors.  From (\ref{he3lim}), the limit
$\eta_{10} > 2.8$ corresponds to the limit $N_\nu < 3.3$ \cite{wssok}.
However, it should be noted that for values of $\eta_{10}$ larger than 2.8,
the central or best-fit value for $N_\nu$ is closer to 2 \cite{osa,kk1,hata2}
and the upper bound is actually found to be 
much smaller with a careful treatment of the
uncertainties, $N_\nu \la 3.1$ \cite{osa,kk1}, though this limit
is relaxed somewhat when the distribution for 
$N_\nu$ is renormalized \cite{osb}.
The range in $\eta_{10}$ of $6.6 \pm 1.4$, corresponds to an even tighter
limit on $N_\nu = 2.0 \pm 0.3$ \cite{hata2} and indicates a problem
when trying to make use of D and \he3 in conjunction with \he4.

Given the magnitude of the problems concerning \he3 \cite{orstv}, 
it would seem unwise to
make any strong conclusion regarding big bang nucleosynthesis 
which is based on \he3.  In addition, deuterium is highly sensitive to models
of chemical evolution. Values of primordial 
D/H as high as $2 \times 10^{-4}$ (\ie\ as
determined in some measurements of D/H in quasar absorption systems
\cite{quas1}) were shown to be viable in models of chemical evolution 
and consistent with other observational constraints \cite{scov}.
For these reasons it was argued \cite{fo,fkot}
that the analysis comparing BBN theory 
and observations should be based primarily on the two isotopes which 
are the least sensitive to the effects of evolution, namely, \he4
and \li7.

In \cite{fkot}, the baryon-to-photon ratio, $\eta$ was determined
on the basis of a likelihood analysis using \he4 and \li7. 
Because of the monotonic dependence of \he4 on $\eta$, the
\he4 likelihood distribution has a single (but relatively broad)
peak at $\eta_{10} = 1.75$. \li7, on the other hand, is not monotonic
in $\eta$, the BBN prediction has  a minimum at $\eta_{10} \simeq 3$
and as a result, for an observationally determined value of \li7 above the
minimum, the \li7 likelihood distribution will show two peaks, in this 
case at $\eta_{10} = 1.80$ and $\eta_{10} = 3.5$. 
The combined likelihood distribution
reflecting both \he4 and \li7 is simply the product of the two individual
distributions.
It was found that when restricting the analysis to the standard model,
including $\nnu = 3$, that the best fit value for $\eta_{10}$ is 1.8
and that
\beq
1.4  <  \eta_{10}  <   4.3  \qquad 95\% {\rm CL}
\label{etar}
\eeq
In determining (\ref{etar}) systematic errors were treated as Gaussian
distributed.
When D/H from quasar absorption systems (those showing a 
high value for D/H \cite{quas1}) is included in the analysis this range is
cut to $1.50 < \eta_{10} < 2.55$.

In \cite{fkot}, the range in $\eta$ from the likelihood analysis 
was translated into a range for $\nnu$. Because of the agreement
between the likelihood predictions of \he4 and \li7 for $\nnu = 3$, the
best fit value was found to be very close to $\nnu = 3$. Since the predicted
\he4 abundance is sensitive to $\nnu$ (and $\eta$), the uncertainties in
in the helium abundance can be translated into an uncertainty in $\nnu$. 
The resulting best-fit to $N_\nu$ based on \he4 and
\li7 was found to be \cite{fkot}
\beq
N_\nu = 3.0 \pm 0.2 \pm 0.4^{+ 0.1}_{- 0.6} 
\label{Nlim2}
\eeq
thus preferring the 
standard model result of $N_\nu = 3$ and leading to $N_\nu < 3.90$ 
at the 95 \% CL level when adding the errors in quadrature.
In (\ref{Nlim2}), the first set of errors are the
statistical uncertainties primarily from the observational
determination of $Y$ and the measured error in the neutrino half-life,
$\tau_n$. The second set of errors is the systematic uncertainty
arising solely from $^4$He, and the last set of errors from the
uncertainty in the value of $\eta$ and is determined by the combined
likelihood functions of \he4 and \li7, \ie\ taken from
Eq.\ (\ref{etar}). A similar result is obtained when the high
D/H value indicated by some observations of quasar absorption systems
is included in the analysis.

In this paper, we follow the approach of \cite{fo} and \cite{fkot} in
constraining the theory 
solely on the basis of the \he4 and \li7 data.  We extend the 
approaches of those two papers by fully exploring the ($\nnu$, $\eta$) 
parameter space of the theory in a self consistent way.  
We also explore the consequences of the QSO
deuterium data.
In the next section we discuss the observational data, and explain our 
choices of ``observed abundances''.  Section 3 covers the likelihood 
functions we use.  We discuss our results in section 4, and draw 
conclusions in section 5.

\section{Observational Data}

The most useful site for obtaining \he4 abundances 
has proven to be H {\sc ii} regions in irregular galaxies.
Such regions have low (and varying) metallicities, and
thus are presumably more primitive than such regions in
our own Galaxy.  Because the \he4 abundance is known for regions of
different metallicity one can trace its evolution as a function 
of metal content, and by extrapolating to zero metallicity
we can estimate the primordial abundance.

There is a considerable amount of data on \he4, O/H, and N/H
in low metallicity extragalactic H {\sc ii} regions \cite{p,evan,iz}.
In fact, there are over 70 such regions observed with metallicities ranging 
from about 2--30\% of solar metallicity. The data for \he4 vs.\ either
O/H or N/H is certainly well correlated and an extensive analysis \cite{osa}
has shown this correlation to be consistent with a linear relation.
While individual determinations of the \he4 mass fraction $Y$
have a fairly large uncertainty ($\Delta Y \ga 0.010$), the large
number of observations 
lead to a {\it statistical} uncertainty
that is in fact quite small \cite{osa,osc}.  Recent calculations \cite{ost3}
based on a set of 62 metal poor (less than 20 \% solar)
extragalactic H {\sc ii} regions give
\beq
\label{eq:he4}
\yp = 0.234 \pm 0.002 \rm (stat.) \pm 0.005 (syst.)
\label{h1}
\eeq
When this set is further restricted to include only the 32 H {\sc ii} with 
oxygen abundances less than 10\% solar it was found that \cite{ost3}
\beq
\yp = 0.230 \pm 0.003 \rm (stat.) \pm 0.005 (syst.)
\label{h2}
\eeq
We will consider both values in our analysis below.

The primordial \li7 abundance is best determined by studies of
the Li content in various stars as a function of metallicity
(in practice, the Fe abundance).  At near solar metallicity, the Li abundance
in stars (in which the effects of depletion are not manifest)
decreases with decreasing metallicity,
dropping to a level an order of magnitude lower
in extremely metal poor Population II halo
stars  with [Fe/H] $\la -1.3$ ([Fe/H] is defined to be the 
$\log_{10}$ of the ratio of Fe/H relative to the solar value for Fe/H).  
At lower values of [Fe/H], the Pop II abundance remains constant 
down to the lowest metallicities measured, (some with [Fe/H] $< -3$ !)
and form the so-called ``Spite plateau" \cite{spite}.
With Li measured for nearly 100 such stars,
the plateau value is well established.   We use the recent results
of \cite{mol} to obtain the \li7 abundance in the plateau
\beq
\label{eq:li7}
y_7 \equiv \frac{\li7}{\rm H} = (1.6 \pm 0.07) \times 10^{-10}
\label{l}
\eeq
where the error is statistical.  Again, if we employ the basic 
chemical evolution conclusion that metals increase linearly with time,
we may infer this value to be indicative of the primordial Li abundance.

One should be aware that there  are considerable systematic uncertainties
in the plateau abundance.  First there are uncertainties that arise even if 
one assumes that the present Li abundance in these stars is
a faithful indication of their initial abundance. The actual \li7 
abundance is dependent on the method of deriving stellar
parameters such as temperature and surface gravity, and 
so a systematic error arises due to uncertainties in
stellar atmosphere models needed to determine abundances.
While some observers try to estimate these uncertainties, 
this is not uniformly the practice.  
To include the effect of these systematics,  we
will introduce the asymmetric error range $\Delta_1 = {}^{+0.4}_{-0.3}$
which covers the range of central values for \li7/H, when different methods
of data reduction are used (see eg. \cite{osc}).
Another source of systematic error in the \li7 abundance arises due to 
uncertainty as to whether the Pop II stars actually {\it have} preserved all 
of their Li.  
While the detection of the more fragile isotope \li6 in two of
these stars may argue against a strong depletion \cite{sfosw},
it is difficult to exclude depletion of the order of a factor of two.
Furthermore there is the possibility that the primordial Li 
has been supplemented, by the time of the Pop II star's 
birth, by a non-primordial component arising from cosmic ray interactions 
in the early Galaxy \cite{wssof}.  While such a contribution cannot 
dominate, it could be at the level of tens of percent. 
We considered the effects of these corrections in \cite{fkot} and will
not pursue them further here.

Finally, there have been several recent reported measurements of 
D/H in high redshift quasar absorption systems. Such measurements are in
principle capable of determining the primordial value for D/H,
and hence $\eta$,
because of the strong and monotonic dependence of D/H on $\eta$.
However, at present, detections of D/H  using quasar absorption systems
indicate both a 
high and  low value of D/H.  As such, it should be cautioned 
that these values may not
turn  out to represent the true primordial value.
The first of these measurements \cite{quas1} indicated a rather high D/H ratio,
D/H $\approx$ (1.9 -- 2.5) $\times 10^{-4}$.  A 
re-observation of the high D/H absorption system has been resolved into 
two components, both yielding high values with an average value \cite{rh1}
\beq
y_2 \equiv {\rm D/H} = (1.9 \pm 0.4) \times 10^{-4}
\label{d}
\eeq
Other 
high D/H ratios were reported in \cite{quas3}. However, there are reported low
values of D/H in other such systems  \cite{quas2} with values D/H $\simeq 2.5
\times 10^{-5}$, significantly lower than the ones quoted above. 

In addition to our analysis based on \he4 and \li7, we will also
consider as in \cite{fkot} the effects of including D/H
in the likelihood analysis on $\eta$ and $\nnu$.  We will present
results which make use of the high D/H measurements.  We will comment
on the implications of the low D/H measurements on our results as well.

\section{Likelihood Functions}

Monte Carlo and likelihood analyses have by now become a common part
of the study of BBN \cite{kr,skm,kk1,kk2,hata1,hata2,fo,fkot}.
Our likelihood analysis follows that of \cite{fkot}, except that we
now consider the additional degree of freedom, $\nnu$.  We begin with
a likelihood function for the predicted value of \yp
\beq
L_{\rm 4,Theory}(Y, \nnu, \eta) =
    {1\over\sqrt{2\pi}\sigma_{Y}(\nnu,\eta)}
    \exp{\left({-(Y-\yp(\nnu,\eta))^{2}\over2\sigma_{Y}^{2}(\nnu,\eta)}\right)}
\eeq
where $\yp(\nnu,\eta)$ and $\sigma_{Y}(\nnu,\eta)$ represent the results 
of the theoretical calculation. 
In \cite{fkot}, we treated the systematic errors in the \he4 
observations as a either a flat distribution, or as a gaussian.
This led to $L_{\rm 4, Obs}(Y)$ 
being represented by the difference between two error functions,
or as a gaussian, respectively.
Here, we follow the latter prescription, combining the statistical and
systematic errors in \yp\ by 
adding them in quadrature.  The resulting form for $L_{\rm 4, Obs}(Y)$ 
is gaussian and can be expressed as
\beq
L_{\rm 4,Obs}(Y) =
    {1\over\sqrt{2\pi}\sigma_{Y0}}
    \exp{\left({-(Y-Y_0)^{2}\over2\sigma_{Y0}^{2}}\right)}
\eeq
where $Y_0$ and $\sigma_{Y0}$ characterize the observed 
distribution and are taken from Eqs. (\ref{h1}) and (\ref{h2}).
 A full likelihood function for \he4 
is then obtained by convolving the two likelihood functions 
representing the theory and observational data on \he4  
\beq
L_{4}(\nnu, \eta) =
    \int dY\, L_{\rm 4,Theory}(Y, \nnu, \eta) L_{\rm 4, Obs}(Y)
\eeq
 This then leads to a complete likelihood function for 
\he4
\beq
L_{4}(\nnu, \eta) =
{1\over\sqrt{2\pi(\sigma_Y^2(\nnu,\eta)+\sigma_{Y0}^2)}}
\exp\left({-(\yp(\nnu,\eta)-Y_0)^2\over 
2(\sigma_Y^2(\nnu,\eta)+\sigma_{Y0}^2)}\right)
\eeq

We can construct likelihood functions for \li7 and D in a similar manner,
again adding statistical and systematic errors in quadrature
\beq
L_{7}(\nnu, \eta) =
{1\over\sqrt{2\pi(\sigma_7^2(\nnu,\eta)+\sigma_{70}^2)}}
\exp\left({-(y_7(\nnu,\eta)-y_{70})^{2}\over 
2(\sigma_7^2(\nnu,\eta)+\sigma_{70}^2)}\right)
\eeq
\beq
L_{2}(\nnu, \eta) =
{1\over\sqrt{2\pi(\sigma_2^2(\nnu,\eta)+\sigma_{20}^2)}}
\exp\left({-(y_2(\nnu,\eta)-y_{20})^{2}\over 
2(\sigma_2^2(\nnu,\eta)+\sigma_{20}^2)}\right).
\eeq
In these expressions, $y_{70}$ and $y_{20}$ are the observed values of the
\li7 and D abundances,  and 
$\sigma_{70}$ and $\sigma_{20}$ their associated uncertainties, while
those quantities which are shown as functions of $(\nnu,\eta)$ are all
derived from our theoretical BBN calculations.

The quantities of interest in constraining the $\nnu$--$\eta$ plane 
are the combined likelihood functions
\beq
L_{47} = L_4\times L_7
\eeq
and
\beq
L_{247} = L_{2}\times L_{47}.
\eeq
Contours of constant $L_{47}$ represent equally likely points in the 
$\nnu$--$\eta$ plane.  Calculating the contour containing 95\% of 
the volume under the $L_{47}$ surface gives us the 95\% likelihood 
region.  If we wish to impose a constraint in the analysis (for 
example, $\nnu\ge3$) we can search for the contour which contains 95\% 
of the volume under $L_{47}$ which satisfies the constraint.
From these contours we can then read off ranges of $\nnu$ and $\eta$.  

\section{Results}

Using the abundances in eqs (\ref{h1},\ref{l}, and \ref{d}) and adding the 
systematic errors to the statistical errors in quadrature we have
\beq
\yp = 0.234 \pm 0.0054
\eeq
\beq
y_{7} = (1.6 \pm 0.36)\times10^{-10}
\eeq
\beq
y_{2} = (1.9 \pm 0.4) \times10^{-4}
\eeq
These abundances are then combined with our theoretical calculations
to produce the likelihood distributions discussed above.

Figure 1 shows a three-dimensional view of both $L_{47}$ and $L_{247}$
(in arbitrary units).
As one can see, $L_{47}$ is double peaked.  This is due to the
minimum in the predicted lithium abundance as a function of $\eta$
as was discussed earlier. The peaks of the distribution as well as the
allowed ranges of $\eta$ and $\nnu$ are more easily discerned in the 
contour plot of Figure 2 which shows the 50\%, 
68\% and 95\% confidence level contours in $L_{47}$ and $L_{247}$.  
The crosses show the location of the 
peaks of the likelihood functions.  Note that
$L_{47}$ peaks at $\nnu=3.0$, $\eta_{10}=1.8$ (in agreement with 
our previous results \cite{fkot}) and at $\nnu=2.3$,
$\eta_{10}=3.6$.  The 95\% confidence level allows the following ranges
in $\eta$ and $\nnu$
\begin{eqnarray}
1.6\le\nnu\le4.0 \nonumber \\
1.3\le\eta_{10}\le 5.0 
\end{eqnarray}
Note however that the ranges in $\eta$ and $\nnu$ are strongly
correlated as is evident in Figure 2.

Since $L_{2}$ picks out a small range of values 
of $\eta$, largely independent of $\nnu$, its effect on $L_{247}$ is 
to eliminate one of the two peaks in $L_{47}$. $L_{247}$ also 
peaks at $\nnu=3.0$,
$\eta_{10}=1.8$. In this case the 95\% contour gives the ranges
\begin{eqnarray}
2.0\le\nnu\le4.1 \nonumber \\
1.4\le\eta_{10}\le 2.6 
\end{eqnarray}
Note that the additional constraint has raised 
the upper limit on $\nnu$ (slightly) though this is counter to intuition. 

These results can be compared with those of \cite{cst2}, who have used 
a Bayesian analysis to examine BBN.  Though they have not performed the full
two-dimensional likelihood analysis presented here, their 
analog of $L_{47}$  peaks at
$\nnu-\Delta Y/0.016\approx2.8$ and 3.9.  Since they have taken a 
significantly larger value for the observed \he4 abundance
($\yp=0.242+\Delta Y\pm0.003$) this corresponds to $\nnu\approx2.3$ 
and 3.4 (using $\Delta Y=-0.008$).  The remaining difference between 
our results is presumably due to the difference in our assumptions 
about the observed \li7 abundance.  Our results for $L_{247}$ also 
appear to be consistent with those of \cite{hata3}.

Since $\nnu=3$ is well within the range covered by the 95\% contour, it
is legitimate to impose the additional constraint $\nnu\ge3$ \cite{osb}.
Figure 3 shows 
the 50\%, 68\% and 95\% confidence level contours under this condition.
The unconstrained contours are also shown (dashed) for comparison.  
This has a rather small effect on the upper limit to $\nnu$.  In the 
case of $L_{47}$ it has increased to $\nnu\le4.1$,
while for $L_{247}$ it has dropped to $\nnu\le4.0$.

Finally, in figure 4 we consider our alternative choice for the \he4
abundance based on the lowest metallicity set of H {\sc ii} regions \cite{ost3}
from eq. (\ref{h2}) which when adding the systematic errors in quadrature
to the statistical errors gives
\beq
\yp = 0.230 \pm 0.0058.
\eeq
The solid curves show the confidence level contours under these
circumstances (the dashed curves show the old contours, for comparison).
As expected, the lower value of \yp\ drives all the contours towards
lower values of $\nnu$.  The peaks occur at $\nnu=2.7$, 
$\eta_{10}=1.7$ and at $\nnu=2.1$, $\eta_{10}=3.4$.
The corresponding ranges are, for $L_{47}$ 
(unconstrained): $1.3\le\nnu\le3.7$, $1.3\le~\eta_{10}~\le 5.0$;
for $L_{247}$ (unconstrained):
$1.8\le\nnu\le3.9$, $1.3\le~\eta_{10}~\le 2.5$.
When the constraint $\nnu \ge 3$ is applied, we find 
for $L_{47}$ ($\nnu\ge3$): $\nnu\le4.0$ and for 
$L_{247}$ ($\nnu\ge3$): $\nnu\le3.9$.

\section{Conclusions}

We have presented a full two-dimensional likelihood analysis
based on big bang nucleosynthesis and the observations of \he4,\li7 and
certain (high) determinations of D/H in quasar absorption systems.
Allowing for full freedom in both the baryon-to-photon ratio, $\eta$, and the
number of light particle degrees of freedom as characterized by the number  
of light, stable neutrinos, $\nnu$, we have confirmed the successful predictions 
of BBN in a limited range in $\eta$ and a range in $\nnu$ which 
not only encompasses the standard model value of $\nnu = 3$, but whose
likelihood is
peaked at or near that value. The largest value of $\nnu$ allowed at the 95\%
CL is found to be 4.0 at a value of $\eta_{10} = 1.8$,
when using only \he4 and \li7
in the analysis.  Because the high values of D/H observed in certain 
quasar absorption systems \cite{quas1,rh1,quas3} 
agree so well with the predictions made when using
\he4 and \li7 this result is only slightly altered when including D/H.
In contrast, had we used the low D/H values found in \cite{quas2}, and
corresponding to a central value of $\eta_{10} \simeq 6.4$,
there would be virtually
no overlap in the likelihood distributions of this value of D/H and 
$L_{47}$ \cite{ko}. Though we would not claim that these results indicate
a problem with the low D/H measurements per se, they do indicate the 
degree to which they are incompatible with the current determinations of
\he4 and \li7.  Until this incompatibility is resolved, no useful
limit on $\nnu$ can be derived with this method when the low D/H 
abundance is used.

\bigskip

{\bf Acknowledgments}

We thank C. Copi, B. Fields, K. Kainulainen,
and T. Walker for useful conversations.
This work was supported in part by 
DOE grant DE-FG02-94ER40823 at Minnesota, and by NASA grant NAG5-2835
at Florida.

%
%
%
%

\begin{figure}[tbp]
	\centering
	\epsfysize=7in \epsfbox{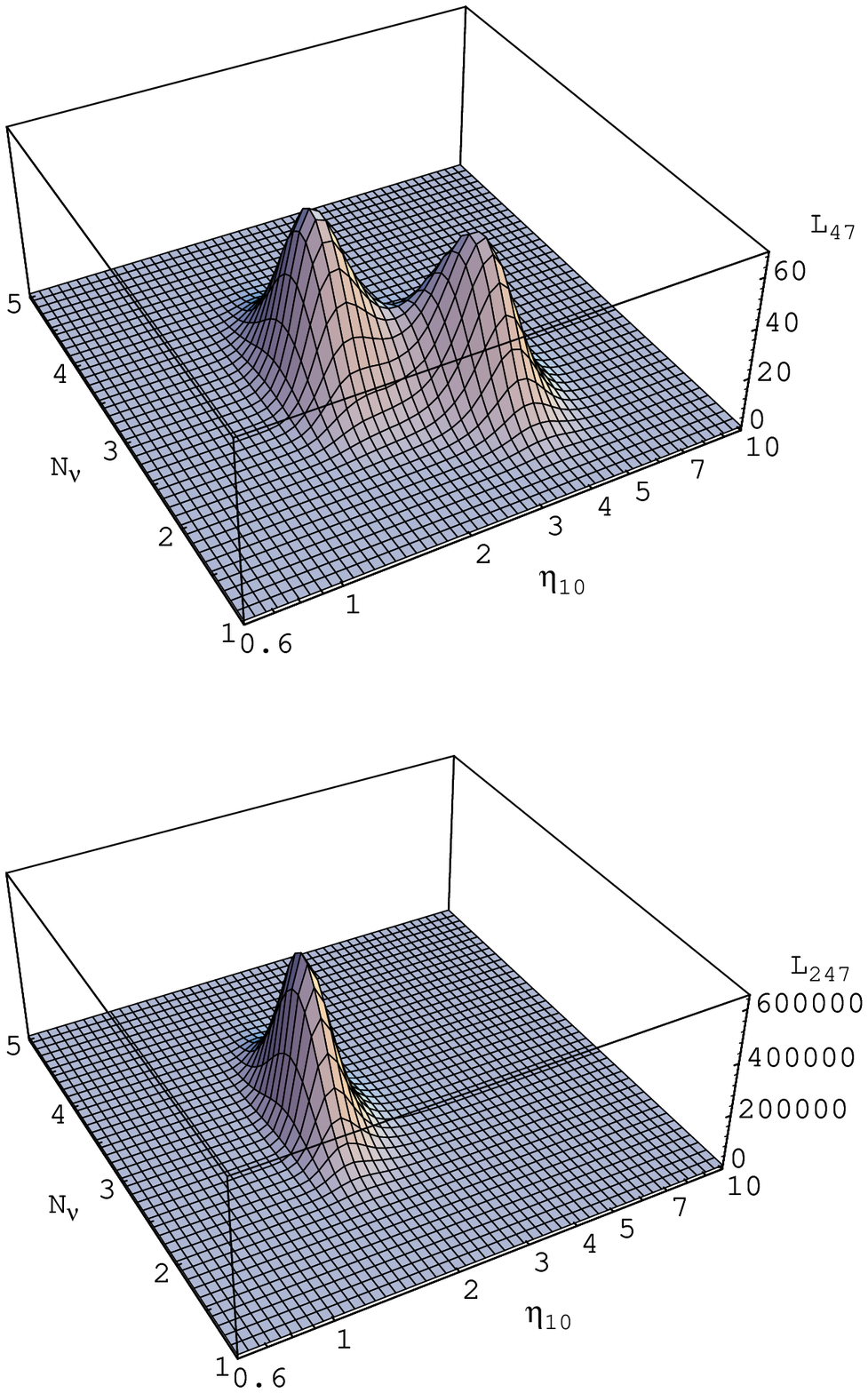}
	\caption{Likelihood functions $L_{47}$ (top) and $L_{247}$ (bottom)
		using $\yp=0.234\pm0.0054$.}
	\label{fig1}
\end{figure}

\begin{figure}[tbp]
	\centering
	\epsfysize=6.4in \epsfbox{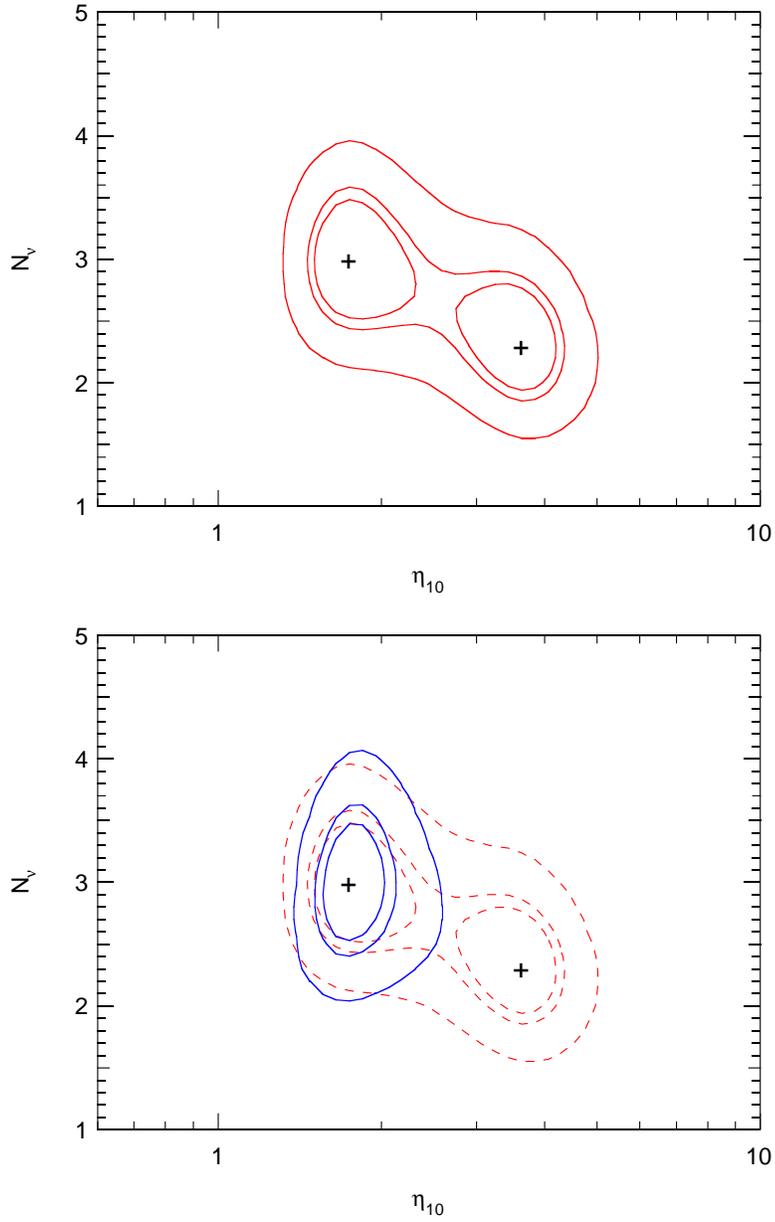}
	\caption{The top panel shows contours in the
		likelihood function $L_{47}$ for $\yp=0.234\pm0.0054$.
		The contours represent 50\% (innermost), 68\% and 95\% 
		(outermost) confidence levels.  The crosses mark the points 
		of maximum likelihood.  The bottom panel shows the 
		equivalent results for $L_{247}$ (with $L_{47}$ shown dashed,
		for comparison.)}
	\label{fig2}
\end{figure}

\begin{figure}[tbp]
	\centering
	\epsfysize=6.4in \epsfbox{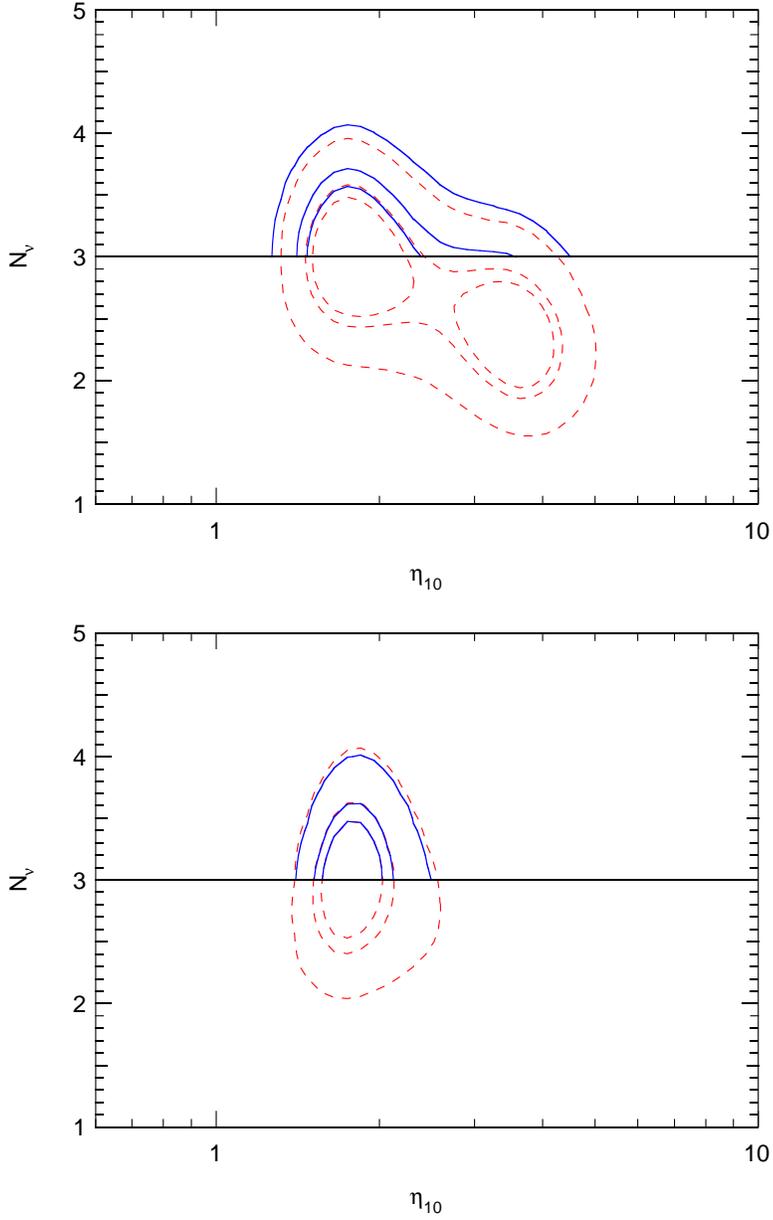}
	\caption{The top panel shows the
		likelihood function $L_{47}$ given the constraint 
		$\nnu\ge3$.  The contours represent 50\%, 68\% and 95\% 
		confidence levels.  The unconstrained likelihood is shown
		dashed.  In both cases $\yp=0.234\pm0.0054$.
		The bottom panel shows the 
		equivalent results for $L_{247}$.}
	\label{fig3}
\end{figure}

\begin{figure}[tbp]
	\centering
	\epsfysize=6.4in \epsfbox{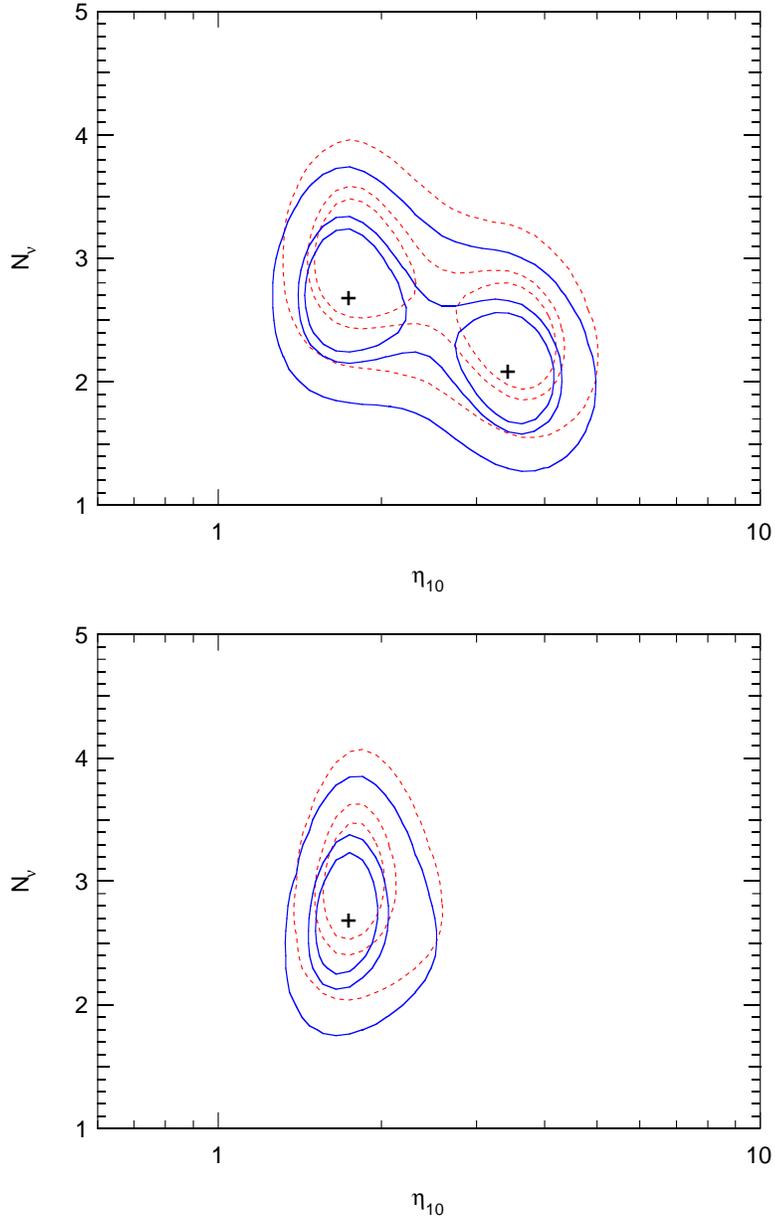}
	\caption{The top panel shows likelihood contours for $L_{47}$,
		with $\yp=0.230\pm0.0058$.  The 
		contours represent 50\%, 68\% and 95\% confidence levels.  
		Results for $\yp=0.234\pm0.0045$ are shown dashed for comparison.
		The bottom panel shows the equivalent results for $L_{247}$.}
	\label{fig4}
\end{figure}

\end{document}